\documentclass{article}
\usepackage{spconf,graphicx}
\usepackage{amsthm}
\usepackage{amsmath}
\usepackage{amssymb}

\usepackage{cite}
\usepackage{graphicx}
\usepackage{psfrag}
\usepackage{subfigure}
\usepackage{url}
\usepackage{stfloats}

\usepackage{float}
\usepackage{bm}

\usepackage{mathrsfs}
\usepackage{subeqnarray}
\usepackage{cases}
\usepackage{stfloats}
\usepackage{subeqnarray}
\usepackage{cases}

\usepackage{algorithm}
\usepackage{algorithmic}

\title{Time-switching Based SWPIT for Network-Coded Two-Way Relay Transmission with Data Rate Fairness}
\name{Ke~Xiong$^\dag$$^\ast$, ~Pingyi~Fan$^\ast$, ~Khaled Ben Letaief$^\ddag$\thanks{This work was supported  by ¡°973¡± program,
no.2012CB316100(2) and the Fundamental Research Funds
for the Central Universities, no.2014JBM024.}}

\address{$^\dag$School of Computer and Information Technology, Beijing Jiaotong University, Beijing, P.R. China\\
$^\ast$Department of Electronic Engineering, Tsinghua University, Beijing, P.R. China\\
$^\ddag$Hong Kong University of Science \& Technology (HKUST), Hongkong}
\begin{document}
%
\maketitle
\begin{abstract}
This paper investigates the simultaneous wireless power and information transfer (SWPIT) for network-coded two-way relay transmission from an information theoretical viewpoint, where two sources exchange information via an energy harvesting relay. By considering the time switching (TS) relay receiver architecture, we present the TS-based two-way relaying (TS-TWR) protocol. In order to explore the system throughput limit with data rate fairness, we formulate an optimization problem under total power constraint. To solve the problem, we first derive some explicit results and then design an efficient algorithm. Numerical results show that with the same total available power, TS-TWR has a certain performance loss compared with conventional non-EH two-way relaying due to the path loss effect on energy transfer, where in relatively low and relatively high SNR regimes, the performance losses are relatively small. \end{abstract}
\section{Introduction}
\label{sec:intro}
As an effective solution for energy harvesting (EH) to prolong the lifetime of energy constrained
wireless systems\cite{R:X1,R:N1}, simultaneous wireless power and information transfer (SWPIT) has recently attract much attention, in which the receiver is able to collect both energy and information from ambient radio frequency (RF) signals\cite{R:X4,R:X5}.
In \cite{R:EP1}, two practical receiver architectures with separated information decoding and energy harvesting, i.e., time switching (TS) and power splitting (PS), were designed for SWPIT and so far they have been investigated for different wireless systems.
Compared with PS, TS is more practical due to its simplicity. Thus, in this paper we consider the TS-based SWPIT for two-way relay channels.

As is known, due to the potential in enhancing network throughput and spectral efficiency, network-coded two-way relaying  has been widely studied over the past decade\cite{R:R1,R:R3}. However, only a few works (see e.g., \cite{R:EP4}) thus far has began to discuss the SWPIT with separate information receiving and energy harvesting for two-way relayings. Thus, this paper focuses on SWPIT for the network coded two-way relay system with decode-and-forward (DF) operation, where two sources with fixed power supply exchange their information through an energy-constrained and SWPIT-enabled relay node.

Different from current existing works, some differences are deserved to be stressed. Firstly, we investigate the SWPIT for network coded two-way relay networks from an information theoretic perspective, where the two-way relay transmission is considered as the combination of a multi-access (MAC) phase and a broadcast (BC) phase. Secondly, our goal is to explore the potential system throughput performance gain of the network-coded two-way relay channel with TS-based relay receiver architecture. That is, for a given available total power of the two-way relay system, we shall answer the question what is performance loss by using SWPIT at the relay compared with the traditional two-way relay channel without EH technology employed. Comparably, in \cite{R:EP4}, it investigated the outage probability and finite-SNR diversity-multiplexing trade-off, where only amplify-and-forward protocol and PS architecture were considered. Thirdly, we are concerned with a popularly discussed two-way transmission scenario (see, e.g., \cite{R:F1,R:F2,R:F3}), where two sources exchange their information with fair data rate constraint.

The contributions of this paper are summarized as follows. 1) We consider the TS-based relaying for the two-way relay channel (TS-TWR) with DF relaying operation, where the
relay spends some time for energy harvesting and the remaining time for information forwarding.  2) We formulate an optimization problem to explore the maximum sum-rate of the system with data rate fairness under total power constraint by jointly optimizing the time switching factor and power assignment between the two sources. 3) Since the problem is non-convex, by theoretical analysis, we first derive some explicit results associated with the optimization problem. Then, we design an efficient algorithm to solve it. 4) We represent extensive numerical results to discuss the performance of our proposed TS-TWR and then get some useful insights.

\vspace{-6pt}
\section{System Model} \label{Sec:SectII}
Consider a two-way relay model without the source-destination direct link. Half-duplex constraint is considered. Thus, a source phase and a relay phase, are involved in completing each around of information exchange between the two sources,
 ${\rm S}_1$ and ${\rm S}_2$, via  a helping relay ${\rm R}$.
In the source phase, ${\rm S}_1$ and ${\rm S}_2$  transmit their signals  to ${\rm R}$. Such a process can be considered as a network coding mode (analogue network coding or physical layer network coding). In the relay phase, $\rm{R}$ firstly applies multiuser detection to  decode the two messages transmitted from the two sources and then re-encodes them into a new message and broadcast the network-coded message to  ${\rm S}_1$ and ${\rm S}_2$.  Once received the signals from $\rm{R}$, ${\rm S}_1$ and ${\rm S}_2$ can extract the desired information by canceling self-interference, because each of them knows its prior transmitted information in the MAC phase.


The quasi-static Rayleigh fading channel model is considered and each channel coefficient $h_i$ (the complex channel coefficients between ${\rm S}_i$ and ${\rm R}$), remains constant during each around of two-way  relaying. The additive noise at each node is described by the independent circular symmetric complex Gaussian random variables with zero mean and unit variance $\mathcal{CN}\thicksim (0,1)$. $H_i\triangleq|h_i|^2$, which represents the channel-to-noise ratio (CNR) over the two links.

As is known,  the capacity region
of the two-way relay channel with DF relaying operation  is \cite{R:R1,R:R2}
\vspace{-3pt}
\begin{flalign}\label{Con:MACBC}
\mathcal{R}=\mathcal{C}_{\textrm{MAC}} (P_1,P_2, h_1, h_2) \cap \mathcal{C}_{\textrm{BC}}(P_r, h_1, h_2),
\end{flalign}
where $\mathcal{C}_{\textrm{MAC}}$ and $\mathcal{C}_{\textrm{BC}}$ denote the capacity region for the MAC and
BC phases, respectively. $P_1$, $P_2$ and $P_r$ denote the transmission power at ${\rm S}_1$, ${\rm S}_2$ and ${\rm R}$, respectively. For $\mathcal{C}_{\textrm{MAC}}$, its expression can be found in \cite{R:R1}.
 For $\mathcal{C}_{\textrm{BC}}$, since network coding is employed, all broadcasted information has to be decoded at both sources and it has a role of a common message in the BC from the relay to ${\rm S}_1$ and ${\rm S}_2$. By using physical layer network coding, $\mathcal{C}_{\textrm{BC}}$ is then given by \cite{R:F3}
\begin{small}
\begin{flalign}\label{Con:BC}
\mathcal{C}_{\textrm{BC}} (P_r, h_1, h_2)= \Big\{({R_1},{R_2}):\left\{ \begin{array}{l}
 {R_1} \le C({P_r}H_2),\\
 {R_2} \le C({P_r}H_1),\\
 \end{array}\right.\Big\},
\end{flalign}
\end{small}
where $C(x)=\frac{1}{2}\log(1+x)$, $R_i$ is the available information transmission rate from ${\rm S}_i$ to ${\rm S}_j$, $i,j\in\{1,2\}$ and $i\neq j$.

Define $\lambda=\frac{R_2}{R_1}$.
For asymmetric applications, e.g., webpage browsing and file downloading, $\lambda\rightarrow 0$ and for symmetric applications, e.g., online games and video conferences, $\lambda \rightarrow 1$. Here, we consider the case $\lambda \rightarrow 1$. That is, two sources exchange information with data rate fairness. Such a scenario was also wildly considered in some existing works, see e.g. \cite{R:F1,R:F2,R:F3}. Moreover, define $\beta=\frac{H_2}{H_1}$, which is used to describe the channel gain difference of $h_1$ and $h_2$.

The intermediate relay is an energy constrained node, which harvests energy from the signals received from ${\rm S}_1$ and ${\rm S}_2$ firstly and then
uses the harvested energy as a source of transmit power to forward the received information
to the two sources. It is assumed that the energy harvesting and information transfer are carried out
for every received block without any constraint on the minimum power level of the received signal \cite{R:EP1}. Similar to some of current existing works (see e.g., \cite{R:EP1,R:EP2,R:EP3}), we also assume that the processing power required by the transmit/receive circuits at the relay is
negligible as compared to the power used for signal transmission. For simplicity, perfect channel
state information of the system are assumed to be known by all the three nodes.

\vspace{-5pt}
\section{Protocol and Optimization Problem}\label{Sec:SectIII}


In TS-TWR, the total time period $T$ for each around of two-way relaying is divided into three parts. The first $T\theta$ $(0 \leq \theta \leq 1)$ is used for the relay to harvest energy from the sources. The remaining block time, $(1-\theta)T$ is used for information transmission, such that half of it, $(1-\theta)T/2$, is used for the source to relay information transmission and the remaining half, $(1-\theta)T/2$, is used for the relay to destination information transmission.
By adopting the energy receiving architecture proposed by \cite{R:EP1}, the energy harvested at $\rm{R}$ in the energy harvesting stage can be given by
$
E_r=(P_1H_1\eta+P_2H_2\eta)\theta T,
$
where $\eta\in(0,1]$ is the energy conversion efficiency which depends on the rectification process and the
energy harvesting circuitry \cite{R:EP1}. It is assumed that all the harvested energy is used in the relay forwarding stage, so the average power for the relay node can be given by
$
P_r=\tfrac{E_r}{T(1-\theta)/2}=(P_1H_1\eta+P_2H_2\eta)\tfrac{2\theta}{1-\theta}.
$
Substituting $P_r$ and the time splitting factor into $\mathcal{C}_{\textrm{MAC}}$ and  (\ref{Con:BC}) and combining them with (\ref{Con:MACBC}), we can express the achievable rate region of TS-TWR as (\ref{Rregion:TS}).
\begin{figure*}[ht!]
\vspace{-20pt}
\begin{flalign}\label{Rregion:TS}
\mathcal{R}_{\textrm{TS-TWR}}=\Bigg\{({R_1},{R_2}):\left\{ \begin{array}{l}
 {R_1} \le \frac{(1-\theta)T}{2}\min\Big\{\log(1+H_1P_1),\log(1+H_2\frac{2\eta\theta(P_1H_1+P_2H_2)}{1-\theta})\Big\},\\
 {R_2} \le \frac{(1-\theta)T}{2}\min\Big\{\log(1+H_2P_2),\log(1+H_1\frac{2\eta\theta(P_1H_1+P_2H_2)}{1-\theta})\Big\}, \\
 {R_1} + {R_2} \le \frac{(1-\theta)T}{2}\log(1+H_1P_1+H_2P_2).
 \end{array}\right.\Bigg\}.
\end{flalign}
\vspace{-20pt}
\end{figure*}

To explore the system potential capacity of TS-TWR with data rate fairness, we formulate an optimization problem as shown in (\ref{Opt:A1}) to jointly optimize the time switching factor $\theta$ and the available power $P_1$ and $P_2$ under the total available power constraint $P_{\rm tot}$.
The objective is to find the joint optimal
$\theta^*$, $P_1^*$ and $P_2^*$ to maximize the system sum rate.
\begin{flalign}\label{Opt:A1}
\mathop {\max }\limits_{\theta,\omega}\,\,\,\, &R_{\textrm{sum}}=\sum\nolimits_{i=1}^{2}R_i\\
\textrm{s.t.} \,\,\,\,\,&(R_1,R_2)\in \mathcal{R}_{\textrm{TS-TWR}},\,\, \lambda=1,\,\,\,\theta \in (0,1)\nonumber\\
&P_1=P_{\rm tot}\omega,\,\,P_2=P_{\rm tot}(1-\omega),\,\,\omega\in (0,1)\nonumber,
\end{flalign}
where $\omega$ is the power allocation factor between ${\rm S}_1$ and ${\rm S}_2$. It can be seen that to optimize $P_1$ and $P_2$ is equal to optimize $\omega$.

\vspace{-6pt}
\section{Optimal Design of TS-TWR}\label{SEC:TSTWR}
In this Section, we discuss how to solve Problem (\ref{Opt:A1}) for TS-TWR.
Although it is difficult to discuss the joint convexity of Problem (\ref{Opt:A1}) in $\theta$ and $\omega$,
we fortunately found that for a given $\theta$, $R_{\rm sum}$ is concave w.r.t $\omega$, and for a given $\omega$, $R_{\rm sum}$ is also concave w.r.t $\theta$. Therefore, we
solve it as follows.

\subsection{Optimal $\omega^*$ of TS-TWR for a given $\theta$}
For a given $\theta$, by the observation of (\ref{Rregion:TS}), we arrive at the following Theorem 1 by theoretical analysis.

\textbf{Theorem 1.} For a given $\theta$, the optimal $\omega^*$ for TS-TWR is
\begin{flalign}
\omega^*=\left\{ \begin{aligned}
 &\tfrac{2H_2\min\{H_1,H_2\}\eta \theta}{H_1(1- \theta) + 2\min\{H_1,H_2\}\eta \theta (H_2-H_1)},\\
 &\quad\quad\textrm{if}\,\,\theta < \tfrac{1}{(4\min\{H_1,H_2\}\eta + 1)}\,\,\textrm{and}\,\,H_2 > H_1\\
 &\tfrac{H_2(1-\theta - 2\min\{H_1,H_2\}\eta \theta)}{H_2 (1-\theta) + 2\min\{H_1,H_2\}\eta \theta (H_1- H_2)},\\
 &\quad\quad\textrm{if}\,\,\theta < \tfrac{1}{(4\min\{H_1,H_2\}\eta + 1)}\,\,\textrm{and}\,\,H_1\geq H_2\\
 &\tfrac{H_2}{H_1 + H_2},\,\,\textrm{otherwise}
 \end{aligned}\right.
\end{flalign}

\begin{proof}
Defining $\Omega=\min\{H_1,H_2\}\frac{2\eta\theta}{1-\theta}$. We have
$
R_1^*=R_2^*\leq K\ast\max\limits_{\omega}\,\,\log\big(1+\min\big\{f_1(\omega),f_2(\omega),f_r(\omega)\big\}\big),\nonumber
$
where $K=\frac{(1-\theta)T}{2}$ $f_1(\omega)=H_1P_{\rm tot}\omega$, $f_2(\omega)=H_2P_{\rm tot}(1-\omega)$ and $f_r(\omega)=\Omega(P_{\rm tot}H_2+P_{\rm tot}\omega (H_1-H_2))$. According to the four cases of the three linear functions w.r.t $\omega$, Theorem 1 can be easily proved.
Theorem 1 is thus proved.
\end{proof}

\subsection{Optimal $\theta^*$ of TS-TWR for a given $\omega$}

For a given $\omega$, both $P_1$ and $P_2$ are determined. 
In this case, we obtain the following results.

\textbf{Lemma 1.} For a given pair of available power $\{P_1, P_2\}$, the optimal rate pair $(R_1^*,R_2^*)$ of Problem (\ref{Opt:A1}) satisfies that
$
R_1^*=R_2^*\leq\mathop {\max }\limits_{\theta} \{K\log(1+\min\{\mathcal{Q},\mathcal{G}\tfrac{2\theta}{1-\theta}\})\},
$
where $\mathcal{G}\triangleq \min\{H_1,H_2\}\eta(P_1H_1+P_2H_2)$ and $\mathcal{Q}\triangleq \min\{H_1P_1,H_2P_2\}$.
\begin{proof}
For a given $\omega$, by the observation of (\ref{Rregion:TS}), it can be easily deduced that the optimal rate pair $(R_1^*,R_2^*)$ satisfies $R_1^*=R_2^*
\leq\max\limits_{\theta}\,\,\tfrac{(1-\theta)T}{2}\log(1+\min\{H_1P_1,H_2P_2,\mathcal{M}\})$.
With $\mathcal{Q}$ and $\mathcal{G}$, one can arrive at Lemma 1.
\end{proof}

 Let $F_1(\theta)=\tfrac{(1-\theta)T\log(1+\mathcal{Q})}{2}$ and $F_2(\theta)=\tfrac{(1-\theta)T\log(1+\tfrac{2\theta\mathcal{G}}{1-\theta})}{2}$. We have the following to lemmas.

\textbf{Lemma 2.} $F_1(\theta)$ is a linearly decreasing function and $F_2(\theta)$ is a firstly increasing and then decreasing function.
\begin{proof}
This Lemma can be proved by deriving $F_2^{'}(\theta)$ and $F_2^{''}(\theta)$. Thus, detail information is omitted here.
\end{proof}

\textbf{Lemma 3.} $F_1(\theta)$ and $F_1(\theta)$ have one and only intersection point.
\begin{proof}
It is known that $0<\theta<1$, so $\frac{(1-\theta^*)T}{2}\neq 0$. In this case, only when $\log(1+\mathcal{Q})=\log(1+\mathcal{G}\frac{2\theta^*}{1-\theta^*})$, $F_1(\theta)=F_2(\theta)$. Hence, it can be deduced that that $\theta^*=\frac{\mathcal{Q}}{1+2\mathcal{G}}$.
\end{proof}

\textbf{Lemma 4.} Let $\theta_1=\mathop {\arg }\nolimits_\theta
\{F_1(\theta)=F_2(\theta)\}$ and $\theta_2=\mathop {\arg }\nolimits_\theta
\{\max F_2(\theta)\}$. Then,  $\theta_2=\tfrac{2\mathcal{G}-\mathcal{W}[\tfrac{2\mathcal{G}-1}{e}]-1}{(\mathcal{W}[\tfrac{2\mathcal{G}-1}{e}]+1)(2\mathcal{G}-1)}$
and $\theta_1=\tfrac{\mathcal{Q}}{1+2\mathcal{G}}$.
where $\mathcal{W}[\cdot]$ is the Lambert $\mathcal{W}$ function.
\begin{proof}
According to Lemma 2 and 3, there are two different cases of the relationship between $F_1(\theta)$ and $F_2(\theta)$. For $\theta_1$, $F_1(\theta_1)=F_2(\theta_2)$.
Therefore, $\theta_1=\frac{\mathcal{Q}}{1+2\mathcal{G}}$. For $\theta_2$, it satisfies that
$
\frac{{d {F_2}}}{{d \theta }}\big|_{\theta_2}=0,
$
so
$
\tfrac{-T}{2}\log(1+\tfrac{2\mathcal{G}\theta_2}{1-\theta_2})+\tfrac{GT}{1-\theta_2+2\mathcal{G}\theta_2}=0.
$
Let $m\triangleq1-\theta_2+2\mathcal{G}\theta_2$. Then,
$\theta_2=\frac{m-1}{2\mathcal{G}-1}$
and
$
\log\big(1+\tfrac{2\mathcal{G}(m-1)}{2\mathcal{G}-m}\big)=\tfrac{2\mathcal{G}}{m}.
$
As a result,
$
e^{\tfrac{2\mathcal{G}}{m}}=\tfrac{(2\mathcal{G}-1)m}{2\mathcal{G}-m}.
$
Thus,
$
m=\tfrac{2\mathcal{G}}{\mathcal{W}[\tfrac{2\mathcal{G}-1}{e}]+1},
$
where $\mathcal{W}[\cdot]$ is the Lambert $\mathcal{W}$ function and $W(z)e^{W(z)}= z$.
Thus, Lemma 4 can be proved.
\end{proof}

\textbf{Theorem 2.} For a given available power pair $\{P_1, P_2\}$, the optimal time switching factor for a TS-TWR system with data rate  fairness is
\vspace{-5pt}
\begin{flalign}\label{Eqn:RTS}
&R_{\textrm{sum}}^{({\tiny \textrm{TS-TWR}})*}=\\
&\left\{ \begin{array}{l}
 \min\big\{(1-\theta_1)\log(1+\mathcal {\mathcal{Q}}),\\
 \quad\quad\tfrac{1-\theta_1}{2}\log\big(1+P_1H_1+P_2H_2\big)\big\},\,\,\textrm{if}\,\,\theta_1\leq \theta_2,\\
 \min\big\{(1-\theta_2)\log\big(1+\mathcal{G}\tfrac{2\theta_2}{1-\theta_2}\big),\\
 \quad\quad\tfrac{1-\theta_2}{2}\log\big(1+P_1H_1+P_2H_2\big)\big\},\,\,\,\textrm{Otherwise}.\\
 \end{array}\right.\nonumber
\end{flalign}
\begin{proof}
From Lemma 1, it is known that
$R_1^*=R_2^*
\leq=\max\limits_{\theta}\min\{F_1(\theta),F_2(\theta)\}.
$
Thus, when $\theta_1 \leq \theta_2$, $\theta^*=\theta_1$ and when $\theta_1 \leq \theta_2$, $\theta^*=\theta_2$. Combining it with Lemma 4 and the MAC joint data rate constraint, we arrive at Theorem 2.
\end{proof}

\subsection{Joint Optimization of $\theta^*$ and $\omega^*$ for TS-TWR}
Base on above results, we design an iterative algorithm to jointly optimize $\theta$ and $\omega$  with a given bias error $\epsilon$.
\begin{algorithm}[h]
\caption{Finding the joint optimal $\{\theta^*,\omega^*\}$}\label{alg:ww}
\begin{algorithmic}[1]
\STATE Initialize $\theta=\tfrac{1}{2}$ and $R_{\textrm{sum}}^{\textrm{(pre)}}=0$;\
\STATE Calculate $\omega$ in terms of Theorem 1;
\STATE Calculate  $R_{\textrm{sum}}^{\textrm{(cur)}}$ in terms of (\ref{Eqn:RTS});\
\label{code:recentStart}
\WHILE {$|R_{\textrm{sum}}^{\textrm{(cur)}}-R_{\textrm{sum}}^{\textrm{(pre)}}|>\epsilon$}
\STATE Update $\theta$ according to Theorem 1;\
\STATE Update $\omega$ in terms of Lemma 4;
\STATE Update $R_{\textrm{sum}}^{\textrm{(cur)}}$ in terms of (\ref{Eqn:RTS});
\ENDWHILE
\STATE Return $\{\theta,\omega\}$.
\label{code:recentEnd}
\end{algorithmic}
\end{algorithm}

Since it can be inferred that $R_{\textrm{sum}}^{\textrm{(TS-TWR)}}$ is concave w.r.t in $\omega$ and $\theta$. Thus,  each round of iteration Algorithm \ref{alg:ww} can improve $R_{\textrm{sum}}^{\textrm{(cur)}}$. As $\theta< 1$ and $\omega< 1$, $R_{\textrm{sum}}^{\textrm{(cur)}}$ cannot be increased without limit. This implies the \textit{convergence} of Algorithm \ref{alg:ww}.
Moreover, it also can be observed that Algorithm \ref{alg:ww} depends on the initialization of $\theta$. In order to reduce the average complexity, we adopt the middle point, i.e., $\tfrac{1}{2}$ for it.

\section{Numerical Results and Discussions}\label{Sec:SectVI}

In this section, we shall discuss the performance of the optimized TS-TWR on the basis of various numerical results. For comparison, we also consider the two-way relay model without EH as a benchmark system. In the non-EH relaying, $P_{\rm tot}$ is optimally allocated to ${\rm S}_1$, ${\rm S}_2$ and $\rm R$ to maximize the system sum rate. According to \cite{R:F2}, for the non-EH two-way relay channel with data fairness,
the maximum sum rate (MSR) is $R_{\textrm{non-EH}}^*=\log(1+\tfrac{2P_{\rm tot}}{V})$, where $V=\tfrac{1}{H_1}+\tfrac{1}{H_2}+\max\{\tfrac{1}{H_1},\tfrac{1}{H_2}\}$.

\begin{figure}
\vspace{-10pt}
\centering
\includegraphics[width=0.7\columnwidth]{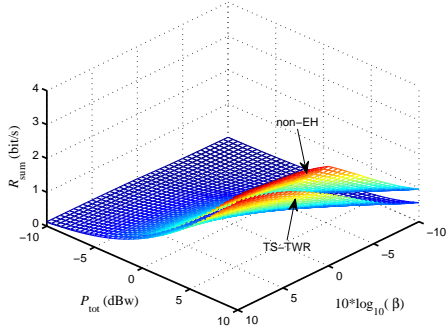}
\vspace{-5pt}
\caption{Comparison of the MSR of TS-TWR and non-EH.}\label{fig:threeface}
\end{figure}

We set $H_1=1$. $10\log_{10}(\beta)$ changing from -10 to 10. $P_{\rm tot}$ is increased from -10dBw to 10dBw. Figure \ref{fig:threeface} shows that
the MSR of the non-EH system is always higher than that of TS-TWR. The reason is that when performing energy transfer, some energy may disperse because of the path loss effect, which may cause loss of system performance compared the non-EH relaying. It also shows that the MSR of TS-TWR increase with the increment of $\beta$ and $P_{\rm tot}$, because the growth of either $\beta$ or $P_{\rm tot}$ can increase the system SNR.

\begin{figure}
\vspace{-5pt}
\centering
\includegraphics[width=0.7\columnwidth]{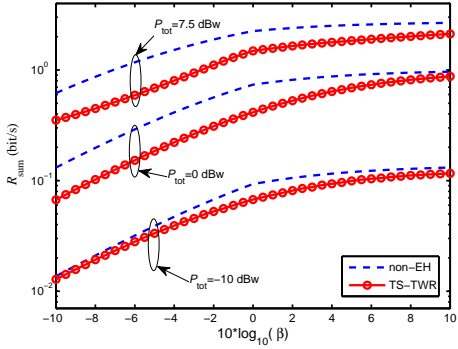}
\vspace{-5pt}
\caption{MSR comparison versus $\beta$.}\label{fig:Threebeta}
\end{figure}

\begin{figure}
\vspace{-10pt}
\centering
\includegraphics[width=0.7\columnwidth]{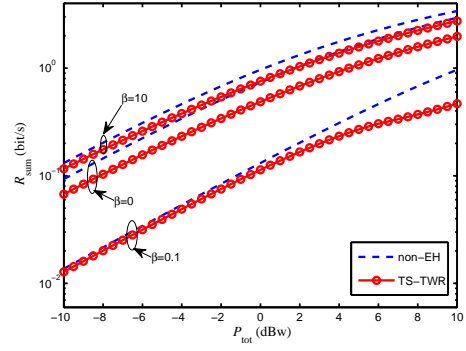}
\vspace{-5pt}
\caption{MSR comparison versus $P_{\rm tot}$.}\label{fig:ThreeP}
\end{figure}

Figure \ref{fig:Threebeta} and Figure \ref{fig:ThreeP} plot the MSR of the two schemes versus $\beta$ and $P_{\rm tot}$, respectively. The two figures shows that the MSR of each scheme increases with the increment of $\beta$ and $P_{\rm tot}$, but the two  schemes show different increasing rate with $\beta$ and $P_{\rm tot}$. Moreover, when $P_{\rm tot}$ is relatively low,  the MSR of TS-TWR is near that of non-EH scheme.

\begin{figure}
\vspace{-5pt}
\centering
\includegraphics[width=0.7\columnwidth]{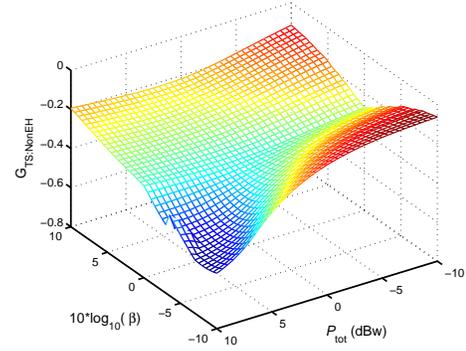}
\vspace{-5pt}
\caption{$G_{\textrm{TS:non-EH}}$ versus $\beta$ and $P_{\rm tot}$.}\label{fig:TS-NON}
\end{figure}

In order to further declare the performance gap between the two shcemes, we define the normalized relative gain of scheme A to scheme B as
$
G_{A:B}=\tfrac{R_{\textrm{sum}}^{({\tiny \textrm{A}})*}-R_{\textrm{sum}}^{({\tiny \textrm{B}})*}}{R_{\textrm{sum}}^{({\tiny \textrm{B}})*}}.
$
$G_{A:B}>0$ implies that scheme A outperforms scheme B while $G_{A:B}<0$, scheme B outperforms scheme A.
It can be seen from Figure \ref{fig:TS-NON} that for TS-TWR, both $\beta$ and $P_{\rm tot}$ have great impacts on its $G_{\textrm{TS:non-EH}}$, where for a relatively small $P_{\rm tot}$, e.g., $P_{\rm tot}=-10$dBw, $G_{\textrm{TS:non-EH}}$ firstly decreases and then increases with the increment of $\beta$ and for a relatively large $P_{\rm tot}$, e.g., $P_{\rm tot}=10$dBw, $G_{\textrm{TS:non-EH}}$ increases with the growth of $\beta$ monotonically.  For a relatively small $\beta$, e.g., $\beta=0.1$, $G_{\textrm{TS:non-EH}}$ monotonically  decreases with the increment of $P_{\rm tot}$ while for a relatively large $\beta$, e.g., $\beta=10$dBw, $G_{\textrm{TS:non-EH}}$ first decreases and then increases with the growth of $P_{\rm tot}$.

\vspace{-5pt}
\section{Conclusion}\label{Sec:SectVII}
This paper studied the SWPIT-aided network-coded two-way relaying, where the relay needs to harvest energy from the wireless transmitted signals from both sources. We considered the TS-TWR from an information theoretical perspective. To explore the system throughput limit under the data fairness constraint, we formulated an optimization problem and derived some explicit results. Numerical results showed that with fixed power supplying of relaying mode and such a work exactly provided some insights for the two-way energy harvesting system design.

\bibliographystyle{IEEEbib}
\bibliography{strings,refs}

\end{document}